\documentclass[twocolumn,twoside,slac_two]{revtex4}
\usepackage{graphicx}
\usepackage{fancyhdr}
\begin{document}

\title{Singlet scalar dark matter effects on Higgs boson driven inflation}

\author{S.T. Love}
\affiliation{Department of Physics,
 Purdue University,  
 West Lafayette, IN 47907-2036, U.S.A.}

\begin{abstract}
A minimal candidate for dark matter is provided by a stable standard model singlet hermitian scalar field. 
The quantum mechanical effects of this singlet are explored in a model where the standard model Higgs boson has a large non-minimal coupling to the Ricci scalar and plays the role of the inflaton. Imposition of the slow roll inflation cosmological constraints restricts the allowed values of the Higgs boson mass, its coupling to the dark matter and the dark matter self-coupling. 
\end{abstract}

\maketitle 

\thispagestyle{fancy}

\newcommand{\be}{\begin{equation}}
\newcommand{\ee}{\end{equation}}
\newcommand{\bea}{\begin{eqnarray}}
\newcommand{\eea}{\end{eqnarray}}
\newcommand{\bwt}{\begin{widetext}}
\newcommand{\ewt}{\end{widetext}}

Slow roll cosmological inflation is an attractive idea \cite{reviews} which, in addition to explaining the large scale flatness, homogeneity and isotropy of the present universe, also accounts for the nearly scale invariant primordial perturbations responsible for structure formation.  Most models of inflation invoke a novel scalar degree of freedom, the inflaton, whose potential must be very flat to accommodate the observed size of said fluctuations. However, there is one model which requires no additional degrees of freedom beyond those already appearing in the standard model. It has been argued \cite{Spokoiny} that a successful model of inflation can be built using the standard model Higgs boson as the inflaton provided one includes a sizeable non-minimal gravitational coupling of the Higgs doublet, $H$, to the gravitational Ricci scalar curvature, $R$. After including the quantum mechanical radiative corrections via the running of the various couplings \cite{DeSimone:2008ei}, the cosmic microwave background (CMB) measurements constrain the range of allowed Higgs boson masses. 

On the other hand, the presence of non-baryonic dark matter requires that there exist degrees of freedom not included in the standard model. A minimal extension is to model \cite{Silveria} the dark matter by a singlet, stable, scalar hermitian field, $S$, which can account for the correct primordial abundance and the lack of direct, indirect and accelerator observation. Here we compute the quantum effects of such scalar dark matter on the Higgs-inflaton effective potential and examine the resultant CMB constraints on the model parameter space \cite{CLLT}. As result of the coupling of the Higgs-inflaton boson to the scalar dark matter, the cosmological restrictions allow for lower Higgs boson masses than is the case of the absence of such couplings. The model action is 
\bea
\Gamma&=&\Gamma_{SM}+\int d^4x \sqrt{-g}[\Lambda +\frac{1}{2}m_{Pl}^2 R+\xi R (H^\dagger H -\frac{v^2}{2})\cr
&&+ \frac{1}{2}\xi_S R S^2+\frac{1}{2}\partial_\mu S g^{\mu\nu}\partial_\nu S -\frac{1}{2}m_S^2 S^2 -\frac{\lambda_S}{4}S^4 \cr
&&-\kappa S^2 (H^\dagger H -\frac{v^2}{2}) ],
\eea
where $\Gamma_{SM}$ is the standard model action minimally coupled to gravity and $m_{Pl}^2 = 1/8\pi G_N \approx (2.4 \times 10^{18})^2 ~{\rm GeV}^2$ is the reduced Planck mass. 
Said action includes all terms through mass dimension four consistent with an unbroken $Z_2$ symmetry under which $S$ is odd hence insuring the scalars' stability. 
The model parameter space has already been somewhat restricted by the dark matter identification. For a dark matter scalar mass, $m_S$, near half the Higgs boson mass, $m_h/2$, the dark matter annihilation process via an intermediate Higgs boson line becomes resonant.  Due to this efficient process, the dark matter will remain in thermodynamic equilibrium to a lower temperature consequently yielding too low a relic density unless the Higgs boson to dark matter coupling $\kappa$ is quite small.  On the other hand, for scalar dark matter masses away from half the Higgs boson mass, either heavier or lighter, there is no such resonance annihilation and thus the Higgs boson-dark matter coupling $\kappa$ must be fairly large, $\kappa > {\cal O}(0.1)$, or else the dark matter would decouple at a sufficiently early stage in the evolution of the universe so that there would be more dark matter present today than what is observed.  Thus for most values of $m_S$ ($m_S$ greater than or less than $m_h/2$), the dark matter abundance calculation favors a higher value of $\kappa$($>{\cal O}(0.1))$.
\begin{figure*}[ht]
$\begin{array}{cc}
\includegraphics[scale=0.55]{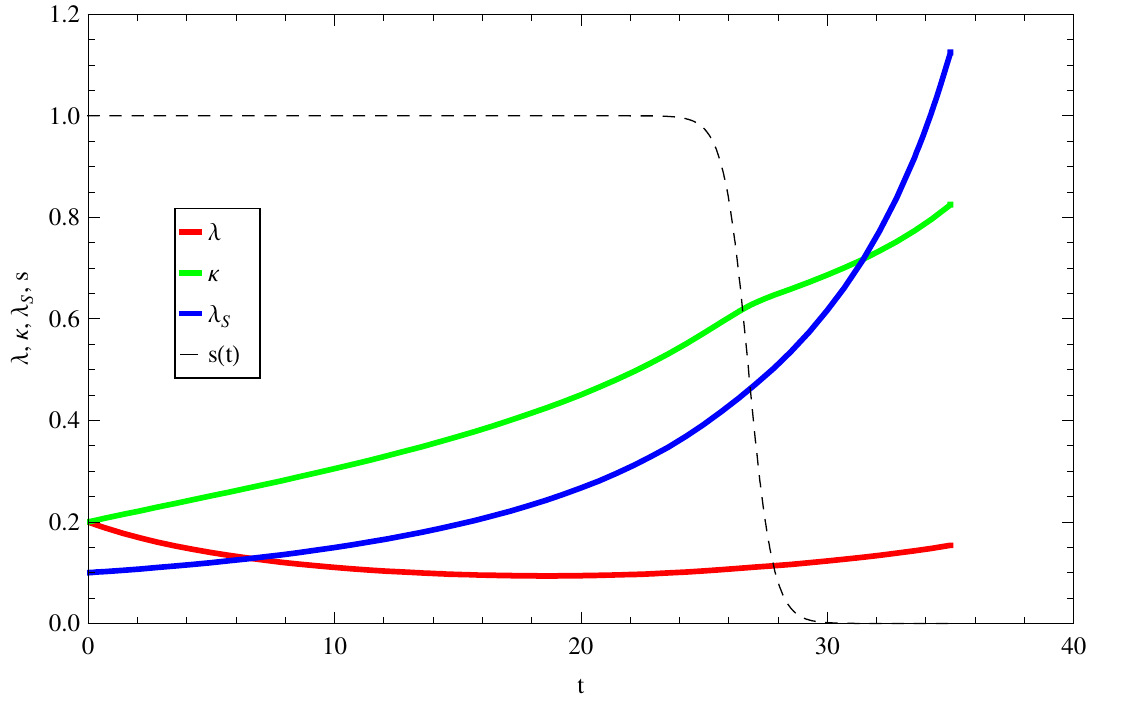}
\includegraphics[scale=0.59]{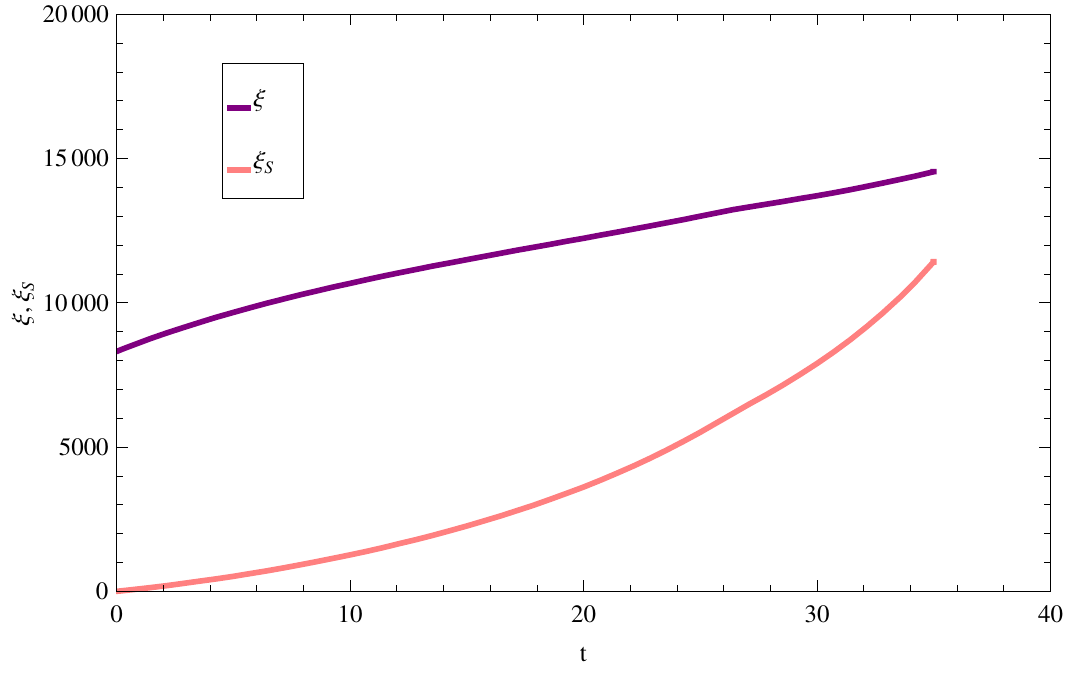}
\end{array}$
\begin{center}
\end{center}
\caption{The running coupling constants for the scalar fields.  The initial conditions for the coupling constants correspond to the effective potential plot of Fig. \ref{EffPotential-1} with $\kappa (0) =0.2$, $\xi (0)=8,315$ and $\xi_S (0) =0.0$.  In this case the onset of inflation occurred at the scale $t_i=35$ with exit at $t_f=32.7$ after 60 e-folds of expansion. Note the rapid fall off of the Higgs propagator modification factor.}
\label{EffCoupling-Suppressed-1}
\end{figure*}
\begin{figure*}
$\begin{array}{cc}
\includegraphics[scale=0.65]{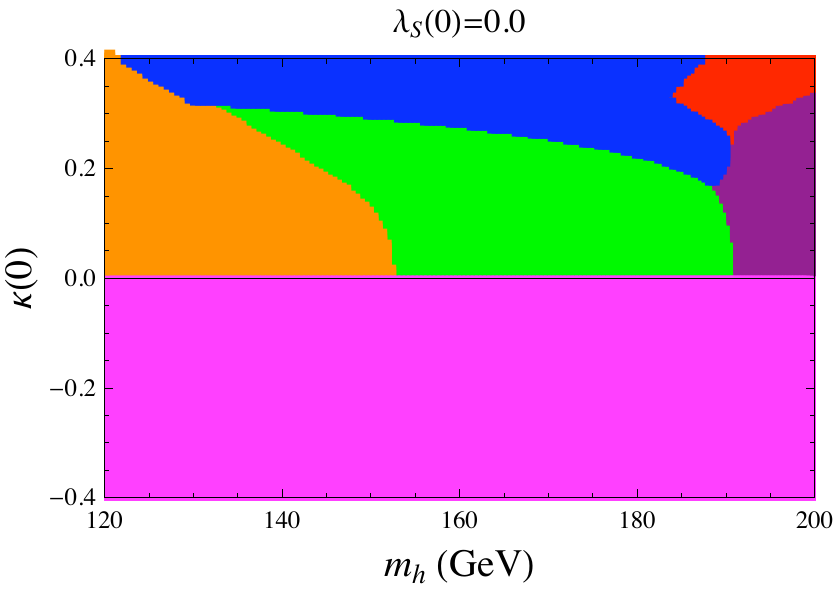} &
\includegraphics[scale=0.65]{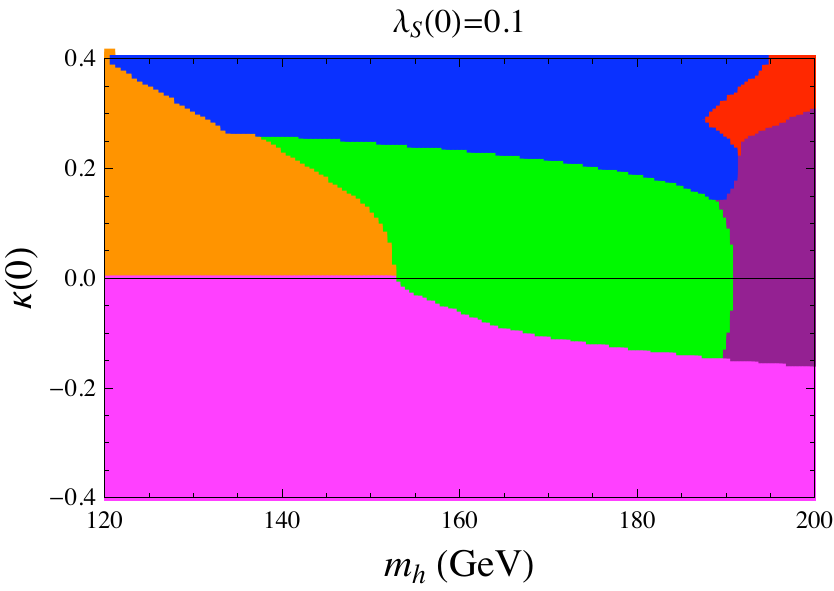} \\
\includegraphics[scale=0.65]{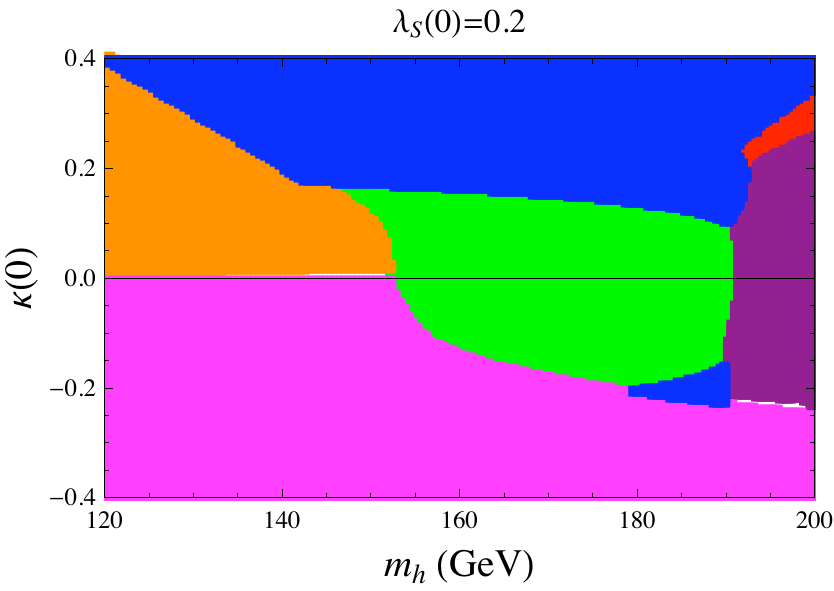} &
\includegraphics[scale=0.65]{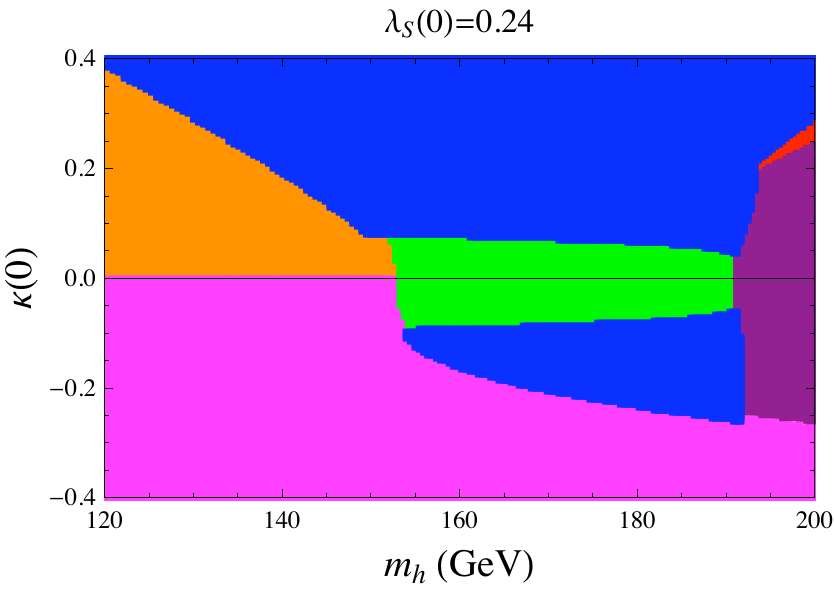} 
\end{array}$
\begin{center}
\includegraphics[scale=0.80]{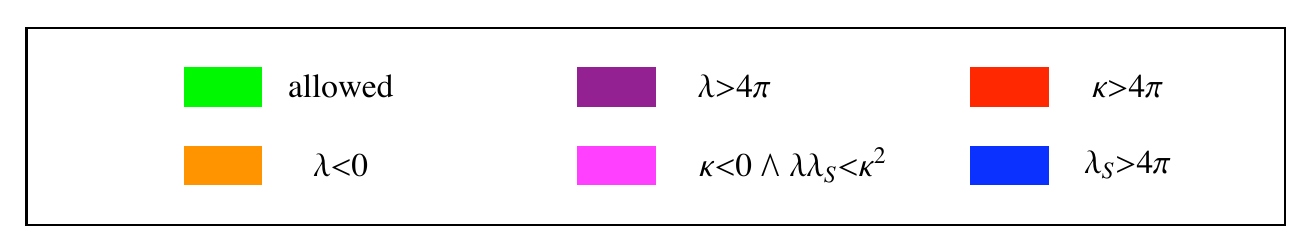} 
\end{center}
\caption{The vacuum stability and triviality bounds on the scalar sector parameter space using $\xi (0) = 10^4$ and including the modified Higgs field propagator. The constraints apply to  $t=34.5$ which is typical of the onset of inflation.  No allowed parameter space remains for $\lambda_S (0) \ge 0.25$.}
\label{suppression}
\end{figure*}

\begin{figure*}
$\begin{array}{cc}
\includegraphics[scale=0.70]{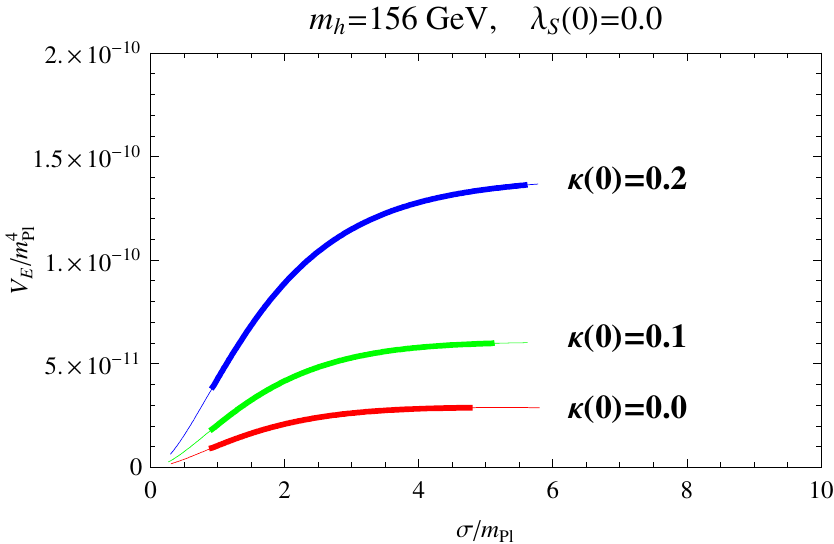}
\includegraphics[scale=0.55]{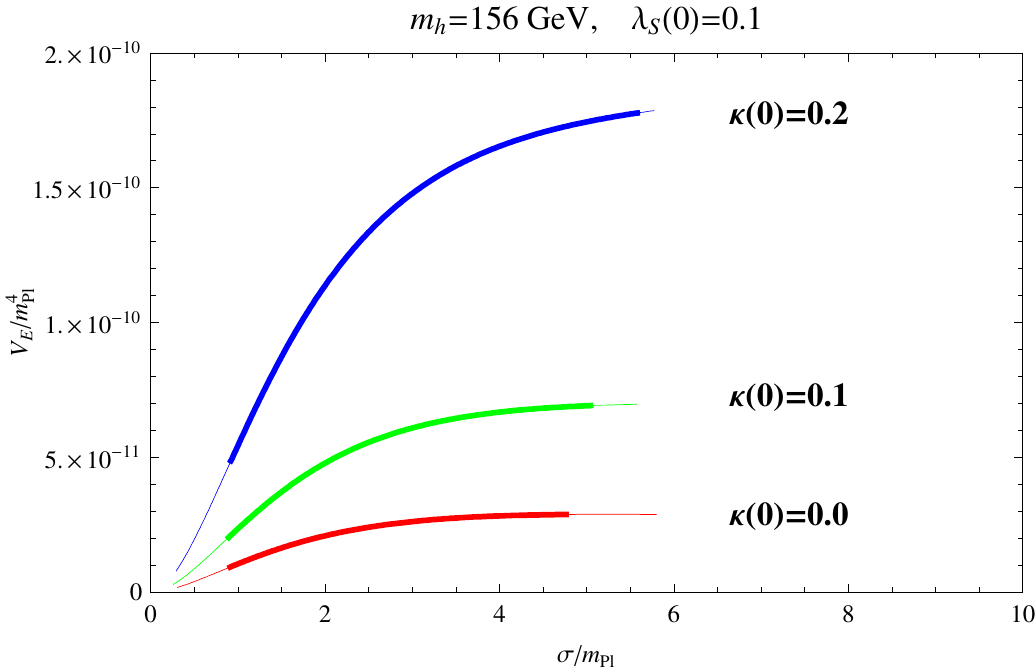}
\end{array}$
\begin{center}
\end{center}
\caption{The Einstein frame renormalization group improved effective potential as a function of the canonically normalized Higgs-inflaton field $\sigma$.  The magnitude and shape of this potential in the inflationary cosmological state varies with the strength of the Higgs-inflaton and dark matter coupling constant $\kappa(0)$.  The thickened portion of the potential curve corresponds to the $N_e =60$ e-folds of inflation with onset and exit values of $\sigma$ as shown. In general, continuing the curves to larger values of $\sigma/M_{Pl}$, the one loop effective potential exhibits a maximum at some point.}
\label{EffPotential-1}
\end{figure*}

The above action is modified by the inclusion of quantum radiative corrections. During the inflationary phase, the physical Higgs-inflaton field has a large expectation value 
$h \sim m_{Pl}/\sqrt{\xi}>>v$, where $v\simeq$ 246 GeV is the scale of electroweak symmetry breaking.
 The mixing of Higgs field with gravity leads to modified Higgs field propagator $
1/p^2 \rightarrow s(h)/p^2  $, 
where 
\bea
s(h)=\frac{1 + \xi h^2 /m_{Pl}^2} {1 +(1+6\xi)\xi h^2 /m_{Pl}^2} .
\label{s(t)}
\eea 
Note that $s(h)\rightarrow 1$ when $\frac{\xi h^2}{m_{Pl}^2}<<1$ while  
$ s(h)\rightarrow \frac{1}{6\xi}$ when $\frac{\xi h^2}{m_{Pl}^2}>>1$. 
Thus internal Higgs field propagation is suppressed for inflationary backgrounds satisfying 
$\frac{\xi h^2}{m_{Pl}^2}>>1$ and $\xi>>1$. After incorporating this single modification, 
the renormalization group functions and improved effective potential for $h=m_te^t$ can be computed using standard techniques. Here we have introduced the scaling variable $t$ and have normalized all fields and couplings at the top quark mass, $m_t$. The running couplings for a particular set of initial conditions and the form of the Higgs propagator modification factor are displayed in Fig, \ref{EffCoupling-Suppressed-1}.
\begin{figure*}
$\begin{array}{cc}
  \includegraphics[scale=0.65]{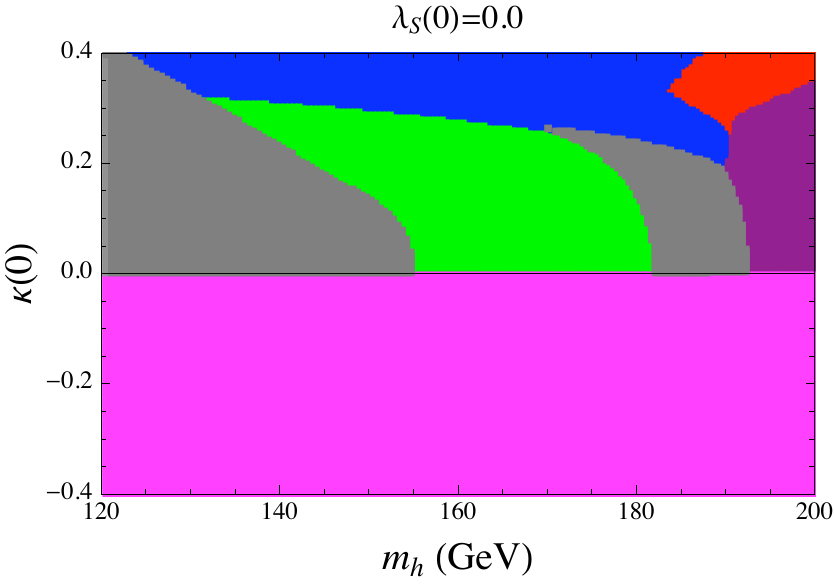}&
  \includegraphics[scale=0.65]{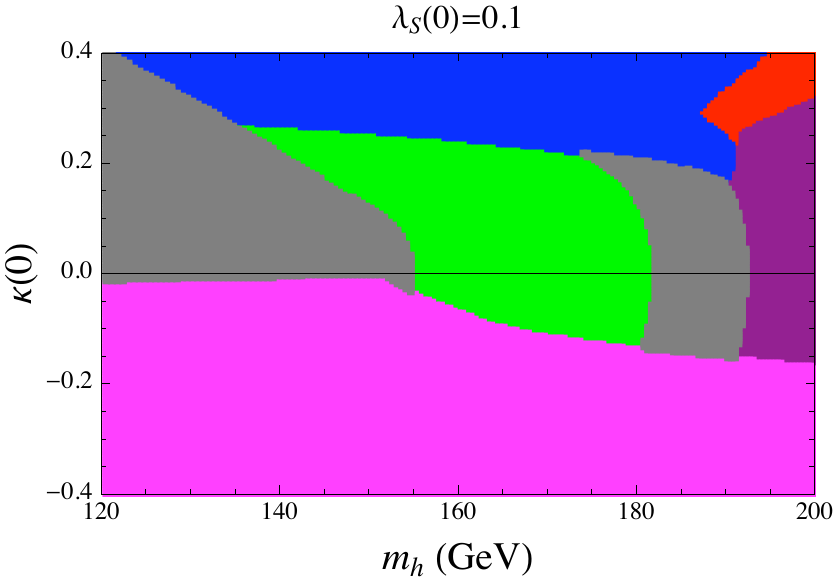}\\
  \includegraphics[scale=0.65]{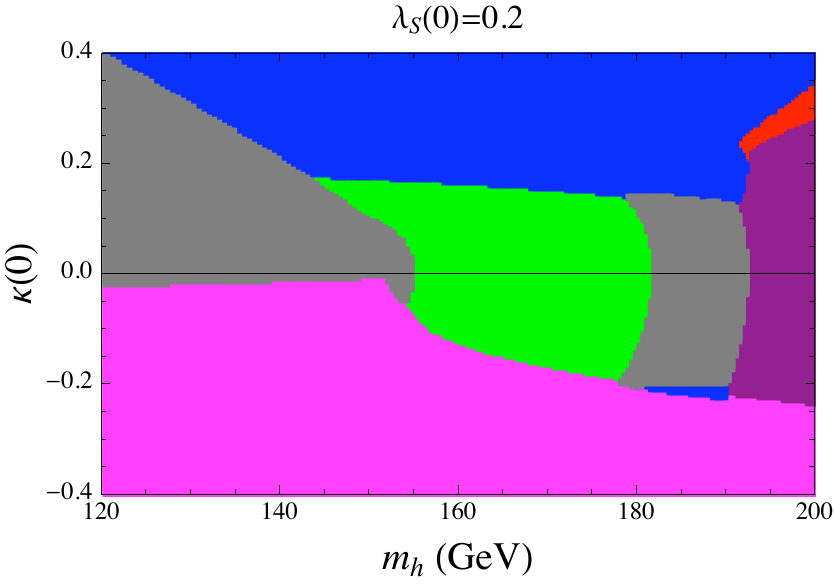}&
  \includegraphics[scale=0.65]{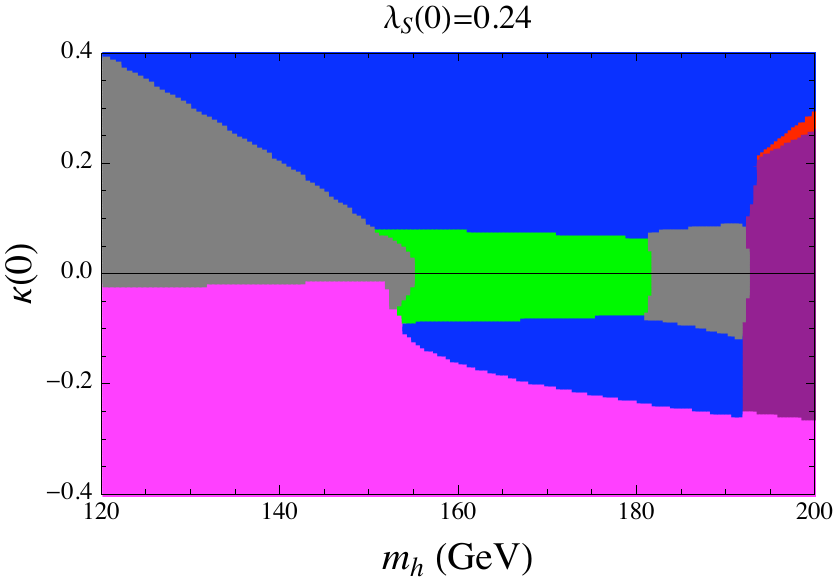}
\end{array}$
\begin{center}
\includegraphics[scale=0.8]{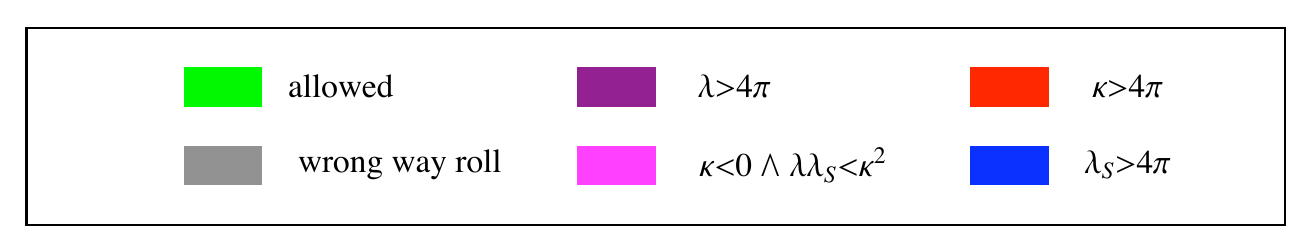} 
\end{center}
\caption{The wrong way roll constraints for parameter space are added to those of vacuum stability and triviality (compare to Fig. \ref{suppression}).  These are displayed for typical initial non-minimal gravitational couplings of $\xi (0) = 10^4$ and $\xi_S (0) =0.0$.  The grey colored areas mark the wrong way roll excluded regions of parameter space.  The constraints apply to scales up to those typical of the onset of inflation, $t_i=34.5$.}
\label{NoRollConstraints}
\end{figure*}
For the model to remain viable, the scalar coupling constants must not reach their respective Landau singularities for all values of $t$ in the range of applicability of the effective theory. We implement these so-called triviality bounds by requiring the coupling constants $\lambda, \kappa, \lambda_S$ to be less than $4\pi$. In addition, vacuum stability requires that, for all values of $t$ in the range of applicability of the effective theory, $\lambda \ge 0$, $\lambda_S \ge 0$ while if $\kappa < 0$ the relation $\lambda~\lambda_S \ge \kappa^2$ must hold. These one loop triviality and vacuum stability bounds including the modified Higgs propagator restrict the allowed parameter space as shown in Fig. \ref{suppression}. Note that smaller values of the Higgs boson mass are allowed with increasing Higgs boson-dark matter coupling $\kappa(0)$ until $\kappa(0)\sim 0.3$ above which the allowed parameter space disappears.

The calculation of cosmological quantities is most conveniently performed in the Einstein frame 
in which the physical Higgs field dependence in the non-minimal gravitational coupling is transformed away. The resultant  Higgs-inflaton effective potential takes the simple form 
\bea
{V}_E = \frac{m_{Pl}^4}{4}\frac{\lambda (t)}{\xi^2 (t)}\frac{\psi^4 (t)}{(1+\psi^2 (t))^2} ,
\eea 
where 
\bea
\psi^2 (t) =\xi (t)  e^{-2\int_0^t dt^\prime \gamma (t^\prime)} e^{2t}m_t^2/m_{Pl}^2 
\eea
is a dimensionless renormalization group invariant and  $\gamma(t)$  the Higgs field anomalous dimension. Fig. \ref{EffPotential-1} displays this effective potential as function of the canonically normalized Higgs-inflaton field $\sigma$ defined as $(\frac{d\sigma}{dt})^2 =m_t^2 \frac{e^{2t}e^{-2\int_0^t dt^\prime \gamma(t^\prime)}}{1+\psi^2(t)}+\frac{3}{2}m_{Pl}^2 \frac{(\frac{d\psi^2(t)}{dt})^2}{(1+\psi^2(t))^2}$. The various cosmological parameters governing the slow roll inflation are then secured in terms of derivatives of the effective potential with respect to $\sigma$ as
\bea
\epsilon &=&\frac{1}{2}m_{Pl}^2 (\frac{1}{{V}_E}\frac{d{V}_E}{d\sigma})^2\cr
\eta &=& m_{Pl}^2 \frac{1}{{V}_E}\frac{d^2{V}_E}{d\sigma^2}\cr
\zeta^2 &=& m_{Pl}^4 (\frac{1}{{V}_E}\frac{d^3{V}_E}{d\sigma^3})(\frac{1}{{V}_E}\frac{d{V}_E}{d\sigma}).
\eea

\begin{figure*}
$\begin{array}{cc}
\includegraphics[scale=0.70]{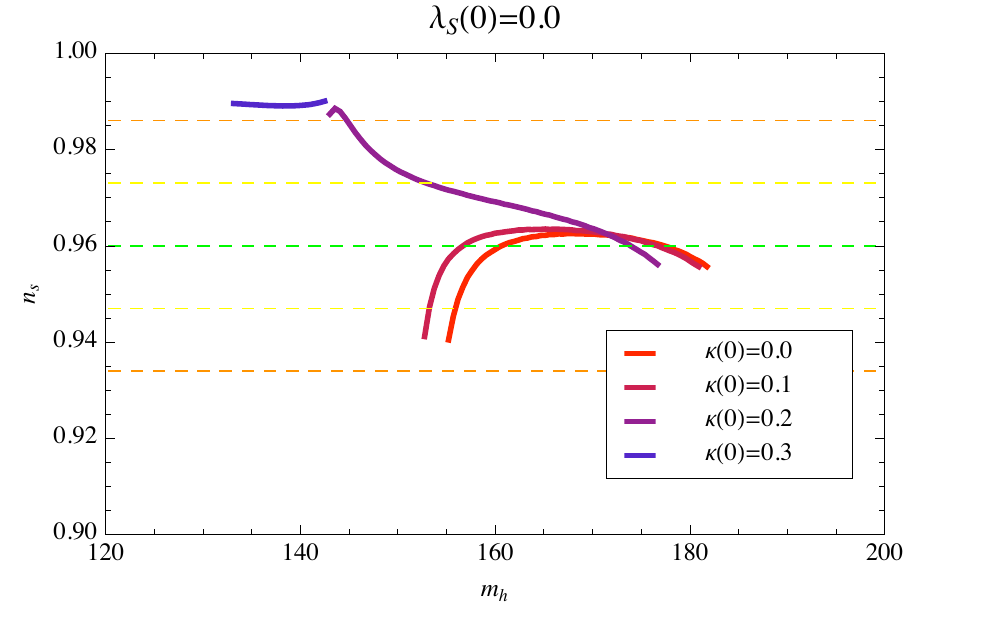} &
\includegraphics[scale=0.70]{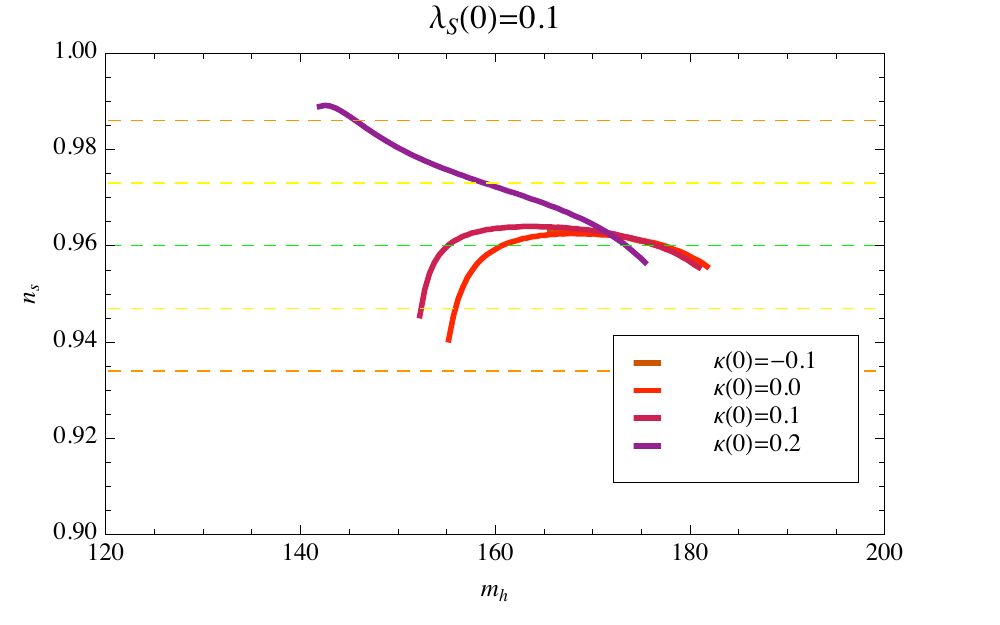} 
\end{array}$
\caption{The spectral index $n_s$, is plotted against the Higgs mass $m_h$ for various values of the Higgs-inflaton to dark matter coupling constant $\kappa (0)$ for the fixed initial value of the dark matter self-couplings $\lambda_S (0) =0.0, 0.1$.  Curve endpoints are determined by the wrong way roll, triviality and vacuum stability conditions. The dashed horizontal lines in the spectral index plot denote its central value, 0.960, and one and two standard deviations from it. Note that the curves corresponding to $\kappa (0) =-0.1$ and $\kappa (0) =0.0$ for the case of $\lambda_S(0)=0.1$ cannot be distinguished up to the thickness of the lines.}
\label{CosmoQ-1}
\end{figure*}
\begin{figure*}
$\begin{array}{cc}
\includegraphics[scale=0.70]{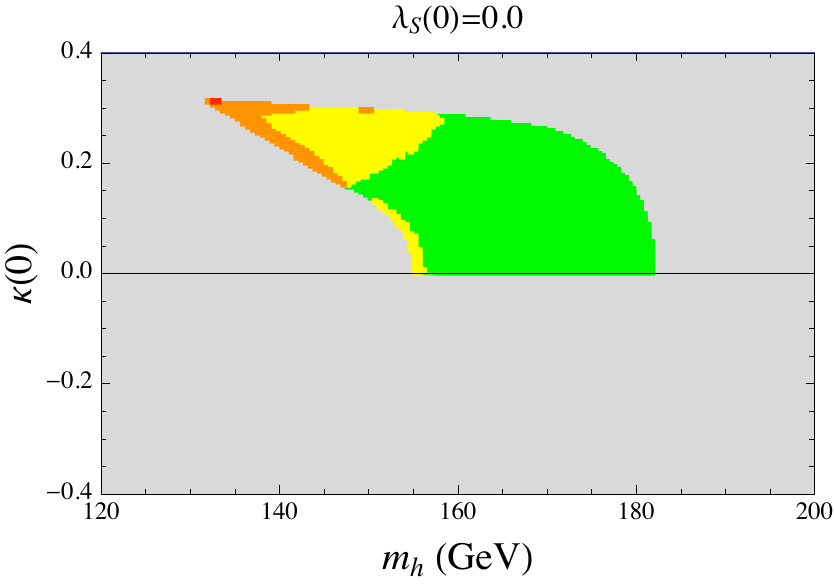}&
  \includegraphics[scale=0.70]{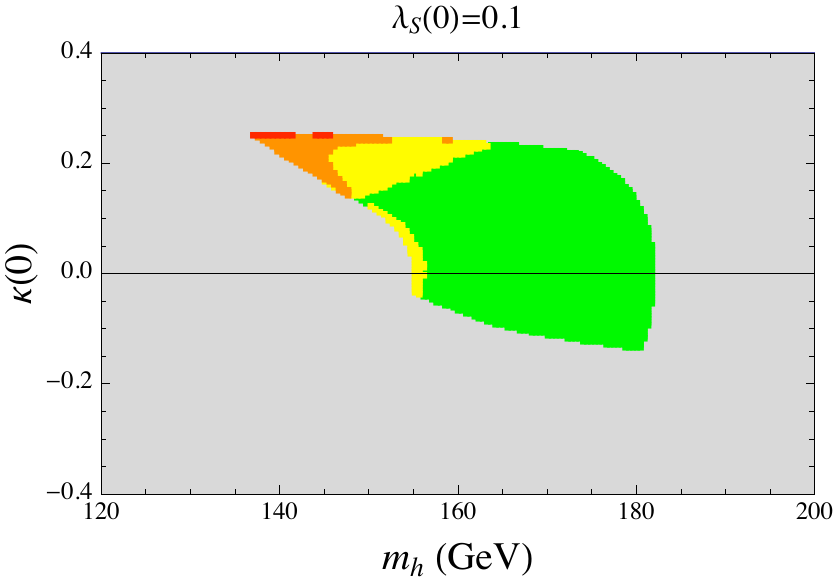}\\
  \includegraphics[scale=0.70]{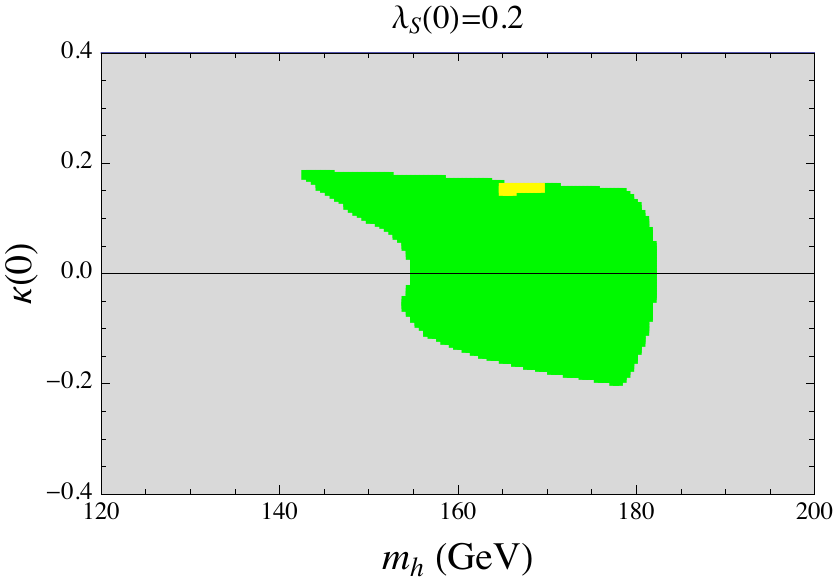}&
  \includegraphics[scale=0.70]{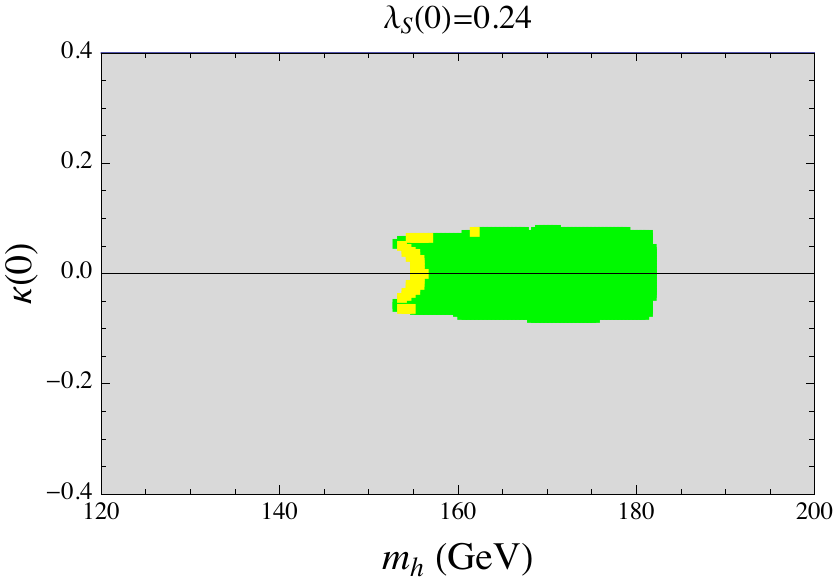}
\end{array}$
\caption{The spectral index cosmological constraints on slices of parameter space for different dark matter self coupling $\lambda_S$ as determined by the degree of agreement with the experimental value of the spectral index, $n_s=0.960$.   Here the green regions indicate the volume of parameter space that predicts spectral index values within one standard deviation of the central value.  The yellow regions correspond to calculated values between one and two standard deviations of the central value, while the orange regions correspond to two to three standard deviations from it.  The red areas indicate parameters that predict spectral index values more than three standard deviations from the central value.  Finally the grey region is excluded by triviality and vacuum stability bounds along with the wrong way roll condition.  For $\lambda_S (0) \ge 0.25$, there is no allowed region of parameter space.}
\label{CosmoParameterSpace-00}
\end{figure*}

In general, the one loop approximated effective potential develops a maximum at some $t$ value as can be gleaned from Fig. \ref{EffPotential-1}. In order to insure that the ensuing rolling will be toward the origin and not toward the Planck mass and beyond, it is necessary that this maximum occur at a smaller $t$ value than the onset of inflation. Imposing this absence of wrong way roll criterion more severely restricts the parameter space of the model than the requirement of absolute vacuum stability up to the onset of inflation. For that matter, even if the absolute stability constraint is abandoned in favor of vacuum meta-stability with a lifetime longer than the age of the observable universe, then the additional range of Higgs boson masses allowed by this less stringent condition are ruled out by the imposition of the wrong way roll constraint.   
For typical values of the scalar parameters the wrong way roll excluded region of parameter space is displayed in Fig. \ref{NoRollConstraints}. For $\kappa(0)=\lambda_S(0)=0$, the allowed range of Higgs boson masses after the imposition of the wrong way roll constraint is roughly 155 GeV $< m_h < $ 182 GeV. This is a slightly smaller range than that allowed without the additional constraint which is 153 GeV $< m_h < $ 190 GeV (c.f. Fig. \ref{suppression}). As $\kappa(0)$ increases, but still with $\lambda_S(0)=0$, the smallest allowed $m_h$ value consistent with the various constraints is approximately 130 GeV, which is roughly the same as without the wrong way roll constraint. This occurs when $\kappa(0)\simeq 0.3$. For larger values of $\kappa(0)$, the allowed parameter space vanishes. As $\lambda_S(0)$ increases from zero, the allowed parameter space starts to shrink as a smaller range of $\kappa(0)$ values are permitted, while there remains a finite range of allowed Higgs boson masses.  Finally, for $\lambda_S(0) >0.25$, the allowed $\kappa(0)$ range vanishes for a finite range of  allowed Higgs boson masses. Hence the allowed parameter space disappears. 

The measured CMB variables provide yet additional restrictions on the allowed parameter space.
\begin{figure*}
$\begin{array}{cc}
\includegraphics[scale=0.70]{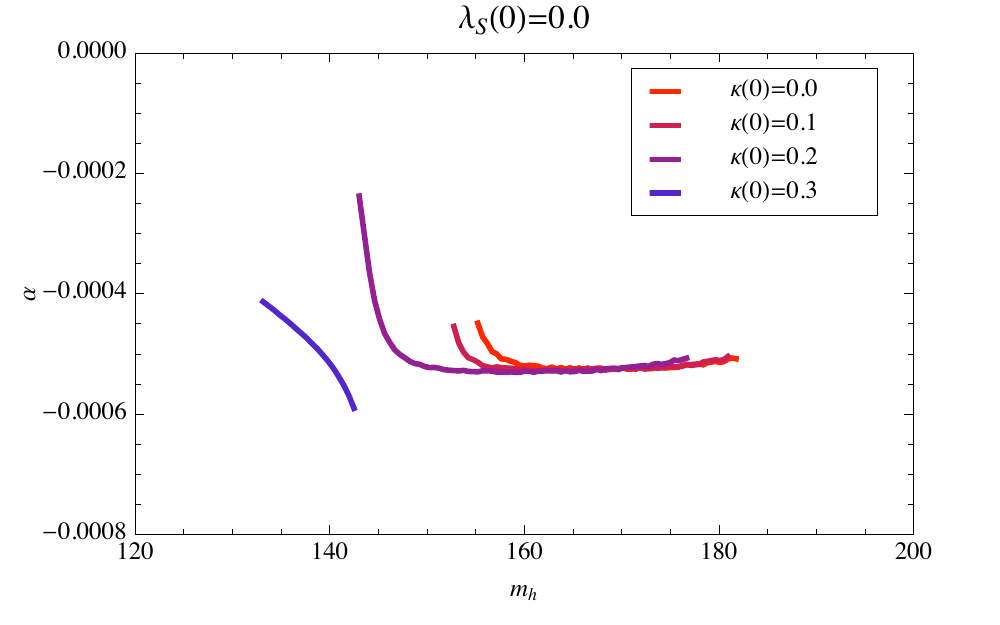}
\includegraphics[scale=0.70]{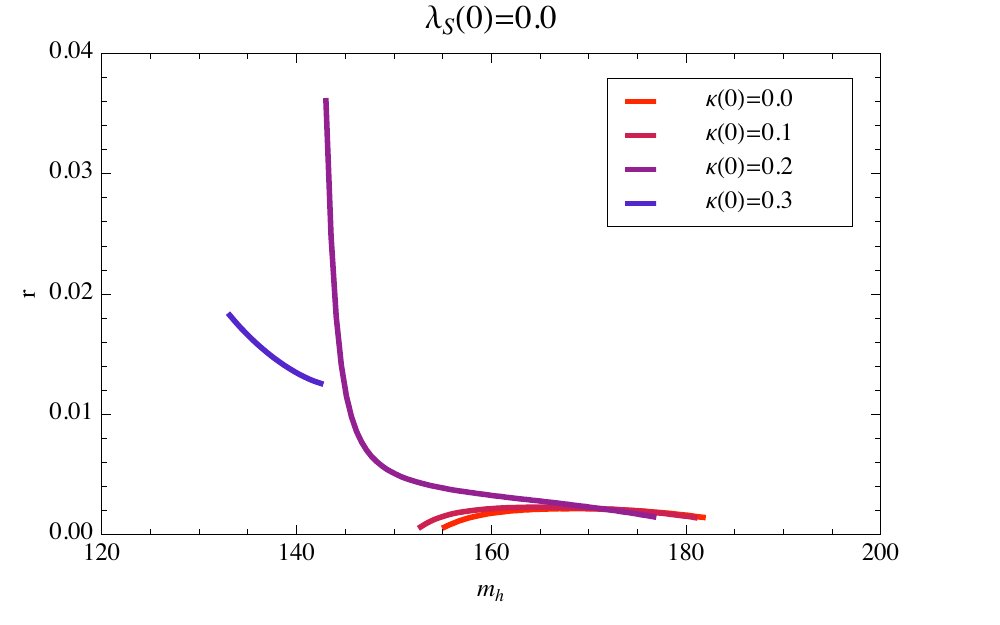}&
\end{array}$
\caption{The spectral index running $\alpha$ and the tensor to scalar ratio $r$ are plotted against the Higgs mass $m_h$ for various values of the Higgs-inflaton to dark matter coupling constant $\kappa (0)$ for the fixed initial value of the dark matter self-coupling $\lambda_S (0) =0.0$.  Curve endpoints are determined by the wrong way roll, triviality and vacuum stability conditions.}
\label{CosmoQ-2}
\end{figure*}
The power spectrum of density fluctuations in $k$-space is given by $
P_s(k)= \Delta^2_{{\cal R}} ({k}/{k^*})^{n_s(k)-1}$ 
where the amplitude of density perturbations is expressed as $
\Delta_{{\cal R}}^2 = \frac{{V_E}}{24 \pi^2 m_{Pl}^4 \epsilon}|_{k^*}$ 
and is secured by the combination of experimental results from WMAP5, BAO and SN to be \cite{Hinshaw:2008kr}: $\Delta^2_{{\cal R}} =(2.445 \pm 0.096)\times 10^{-9}$  at $k^*=0.002$ Mpc$^{-1}$. Slow roll inflation gives the spectral index, $n_s$, its running, $\alpha =dn_s/d\ln{k}$, and the tensor to scalar ratio, $r$, as 
\bea
n_s &=& 1-6\epsilon + 2 \eta \cr
\alpha &=&-24 \epsilon^2 +16 \epsilon \eta -2 \zeta^2 \cr
r &=& 16 \epsilon .
\eea 
The combined  WMAP5, BAO and SN observations \cite{Hinshaw:2008kr} yield 
$\alpha =-0.028\pm 0.020$ (and hence an approximately $k$ independent spectral index ) while $n_s =0.960 \pm 0.013$  and 
$r<0.22$ (95\% CL).

The calculational recipe employed in extracting these parameters 
involves first solving the renormalization group equations with the experimentally determined inputs for $m_t$ and the gauge couplings while scanning over values for $m_h=\sqrt{2\lambda(0)}v, \kappa(0), \lambda_S(0), \xi_s(0)$. The measured $\Delta_R^2$ fixes $\xi(t_i)$ where $t_i$ is the onset of inflation. The exit of inflation, $t_f$, is fixed by the condition $\epsilon(t_f)=1$. The number of e-folds $
N_e(t_i)=\frac{1}{\sqrt{2}m_{Pl}}\int_{\sigma_i}^{\sigma_f} \frac{d\sigma^\prime}{\sqrt{\epsilon(\sigma^\prime)}}$  between $t_i$ and $t_f$ is taken to be 60 and thus fixes $t_i$. The various slow roll cosmological parameters are then evaluated at $t_i$, the determined value of the onset of inflation. In general, as the Higgs boson to dark matter coupling strength, $\kappa (0)$, increases, lower values of the Higgs mass will still support a stable vacuum as well as avoid the wrong way roll condition.  However, a tension begins to arise between the experimentally allowed values of the spectral index and the model predictions of the index at low Higgs boson mass.  It follows from 
Figs. \ref{CosmoQ-1} and  \ref{CosmoParameterSpace-00} 
that, consistent with the triviality, vacuum stability and wrong way roll constraints, agreement with the central measured value of $n_s$ favors a Higgs boson mass in the range 155-180 GeV and a smaller value of $\kappa(0)$. As $\kappa(0)$  grows to values $\sim 0.3$, above which the allowed parameter space vanishes,  the computed value of $n_s$ lies between one to three standard deviations above the central measured value of $0.960$ and occurs for smaller Higgs boson masses of order 130-145 GeV. Thus a discovery of a Higgs boson mass in this range favors both a larger spectral index and a larger coupling of the Higgs boson to dark matter as is also preferred by the dark matter abundance calculations. Finally, for $\lambda_S(0) >0.25$, the allowed $\kappa(0)$ range vanishes for a finite range of  allowed Higgs boson masses and hence the allowed parameter space disappears. Note that there is no additional constraint arising from  $r$ and $\alpha$ as their computed values lie well below the present experimental limits as seen in Fig. \ref{CosmoQ-2}. 

The influence of the inclusion of dark matter on slow roll inflation models where the inflaton is identified with the standard model Higgs boson was explored.  To achieve this, the standard model was modified by the inclusion of an  hermitian scalar standard model singlet field, which can account for the observed abundance of dark matter, and a large non-minimal coupling of the Higgs doublet to the Ricci scalar curvature.  In the inflationary region where the physical Higgs-inflaton field develops a sizeable classical background, the presence of the large Higgs doublet non-minimal gravitational coupling results in a highly suppressed  physical Higgs field propagator. Accounting for this, the one-loop renormalization group improved effective potential was computed and the constraints on the model parameter space were delineated. In addition to the usual triviality and vacuum stability bounds,  focus was given to the cosmological constraints arising from the identification of the Higgs boson with the inflaton. Since, in general, the one loop effective potential develops a maximum, it was necessary to insure that the onset of inflation occurred such that the inflaton rolling was toward the origin and not towards the Planck scale. This wrong way roll constraint was seen to eliminate even more of the parameter space than the vacuum stability (or meta-stability) constraint. Various CMB parameters characterizing the slow roll inflation were computed with the spectral index, $n_s$, providing the most stringent constraint on the coupling constant space. The region of parameter space allowed after the imposition of wrong way roll, triviality and vacuum stability constraints, is further partitioned into various sections whose agreement with the measured spectral index is only at a varying number (one - three) of standard deviations above the central value (c.f. Fig. \ref{CosmoParameterSpace-00}).  Larger values of the coupling of the Higgs-inflaton to dark matter, as is preferred by the dark matter abundance calculation, lead to a lower allowed range of Higgs boson masses which is still consistent with the present accelerator bounds. Thus even after the inclusion of the dark matter, there still remains a range of LHC attainable Higgs boson mass values that are consistent with the cosmological parameters of slow roll inflation when the Higgs scalar is identified as the inflaton.

\begin{acknowledgments}
 This work was supported in part by the U.S. Department of Energy under grant DE-FG02-91ER40681 (Theory and Phenomenology Task). I thank T.E. Clark, B. Liu  and T. ter Veldhuis for an enjoyable collaboration.
\end{acknowledgments}


\bigskip 
\begin{thebibliography}{99}

\bibitem{reviews}For reviews, see, for example, A. Linde, Lect. Notes Phys. {\bf 738}, 1 (2008); 
V. Mukhanov, {\it Physical Foundation of Cosmology} (Cambridge UK Univ. Press, 2005) 421.

\bibitem{Spokoiny}B.L. Spokoiny, Phys.\ Lett.\ B {\bf 147}, 29 (1984); 
D.~S.~Salopek, J.~R.~Bond and J.~M.~Bardeen,  Phys.\ Rev.\  D {\bf 40}, 1753 (1989); 
R. Fakir and W.G. Unruh, Phys. Rev. {\bf D41}, 1753 (1990);
D.I. Kaiser, Phys. Rev. {\bf D52}, 4295 (1995) [arXiv:astro-ph/9408044]; 
E. Komatsu and T. Futamase, Phys. Rev. {\bf D59}, 064029 (1999) [arXiv:astro-ph/99011279]; 
S. Tsujikawa and B. Gumjudpai, Phys. RTev. {\bf D69}, 123523 (2004) [arXiv:astro-ph/0402185]; 
A.O. Barvinsky and A.Y. Kamenshchik, Phys. Lett. B {\bf 332}, 270 (1994) [arXiv:gr-qc/940462]; Nucl. Phys. B {\bf 532}, 339 (1998) [arXiv:hep-th/9803052]; 
  F.~L.~Bezrukov and M.~Shaposhnikov,
  Phys.\ Lett.\  B {\bf 659}, 703 (2008), [arXiv:hep-th/0710.3755];
  A.~O.~Barvinsky, A.~Y.~Kamenshchik and A.~A.~Starobinsky,
  JCAP {\bf 0811}, 021 (2008)
  [arXiv:hep-ph/0809.2104 ];
  F.~Bezrukov, D.~Gorbunov and M.~Shaposhnikov,
  arXiv:hep-ph/0812.3622 .
 


\bibitem{DeSimone:2008ei}
  A.~De Simone, M.~P.~Hertzberg and F.~Wilczek,
  Phys.\ Lett.\ B {\bf 678}, 1 (2009), 
 [arXiv:hep-ph/0812.4946 ];
  F.~L.~Bezrukov, A.~Magnin and M.~Shaposhnikov,
  arXiv:hep-ph/0812.4950 ;
  F.~Bezrukov and M.~Shaposhnikov,
 arXiv:hep-ph/0904.1537 ;
  A.~O.~Barvinsky, A.~Y.~Kamenshchik, C.~Kiefer, A.~A.~Starobinsky and C.~Steinwachs,
  arXiv:hep-ph/0904.1698.



\bibitem{Silveria}
V. Silveria and A. Zee, 
Phys. Lett.  {\bf B161}, 136 (1985); 
  J.~McDonald,
  Phys.\ Rev.\  D {\bf 50}, 3637 (1994)
  [arXiv:hep-ph/0702143].
  C.~P.~Burgess, M.~Pospelov and T.~ter Veldhuis,
  Nucl.\ Phys.\  B {\bf 619}, 709 (2001)
  [arXiv:hep-ph/0011335];
  H.~Davoudiasl, R.~Kitano, T.~Li and H.~Murayama,
  Phys.\ Lett.\  B {\bf 609}, 117 (2005)
  [arXiv:hep-ph/0405097].

\bibitem{CLLT}
T.E. Clark, B. Liu, S.T. Love and  T. ter Veldhuis, arXiv:hep-ph/0906.5595.


\bibitem{Hinshaw:2008kr}
  G.~Hinshaw {\it et al.}  [WMAP Collaboration],
  Astrophys.\ J.\ Suppl.\  {\bf 180}, 225 (2009)
  [arXiv:astro-ph/0803.0732].


\end{thebibliography}
\end{document}